\documentclass[12pt]{iopart}
\usepackage{graphicx}
\usepackage{epsfig}
\usepackage{bm}
\usepackage[flushleft]{threeparttable}
\newcommand{\beq}{\begin{eqnarray}}
\newcommand{\eeq}{\end{eqnarray}}  
\begin{document}

\title[]{Applied electric field on zigzag graphene nanoribbons: reduction of spin stiffness and appearance of spiral spin density waves}

\author{Teguh Budi Prayitno$^1$, Esmar Budi}
 
\address{Department of Physics, Faculty of Mathematics and Natural Science, Universitas Negeri Jakarta, Kampus A Jl. Rawamangun Muka, Jakarta Timur 13220, Indonesia}
\ead{$^1$teguh-budi@unj.ac.id}
\vspace{10pt}

\begin{abstract}
We investigated the reduction of the spin stiffness and the appearance of the spiral spin density waves when the electric field is applied on the zigzag graphene nanoribbons for the ferromagnetic and antiferromagnetic edge states. For that purpose, we exploited the generalized Bloch theorem combined with a constraint method to keep the direction of the magnetic moment of the carbon atom at the edges. We found that the ground state of ferromagnetic configuration is unstable and leads to the spiral ground state while the ground state of the antiferromagnetic configuration is robust. We also showed that the spin stiffness in the ferromagnetic and antiferromagnetic configurations reduces as the electric field increases. Thus, we justified that not only the spin stiffness but also the ground state of the zigzag graphene nanoribbons can be controlled by the electric field.           
\end{abstract}

\vspace{2pc}

\section{Introduction}
The series of experiments on graphene \cite{Novoselov1, Novoselov2, Geim}, a two-dimensional material composed of carbon atoms, has triggered a great enthusiasm for the exploration in nanotechnology. One of the peculiar properties in graphene is the existence of a flat band energy at the Fermi level, which leads to the high density of electron. Recent experiments show some extraordinary behaviors \cite{Liao, Bonac, Schwierz}, such as optical properties and transport properties, as the physical or chemical treatments are employed. At the same time, the theoretical study of the reduction of dimensionality of graphene to become the graphene nanoribbon \cite{Fujita} suggests that graphene and its derivation can be considered as potential candidates for the spintronic devices. 
        
In general, graphene is a kind of nonmagnetic materials, so it is important to consider a magnetism scenario in graphene for the spintronic applications. A simple way to study a magnetism in graphene materials is to lower the dimensionality of graphene. The zigzag graphene nanoribbon (ZGNR), a one-dimensional version of the graphene material, shows many interesting magnetic properties. The magnetism in the ZGNR appears when the magnetic moments of the carbon atoms are arranged at the different edges \cite{Sawada1}. Later, several efforts have been made to manipulate the band gap for future applications in spintronic devices. Among them is the application of external electric field \cite{Rudberg, Kan, Kumar1}, by which either a half-metallic or an insulating character can be induced in the ZGNR.

Here, we focus on the spin stiffness in the ZGNR, which is a crucial indicator of the performance of ZGNR-based applications at the room temperature. Yazyev and Katsnelson showed that the critical temperature of ZGNR should be very low below the room temperature \cite{Yazyev}, which leads to the impossibility of operating ZGNR-based spintronic devices at the room temperature. However, they claimed that a high spin stiffness in the ZGNR is required to drive a long spin-correlation length to increase the critical temperature up to the room temperature. Unlike the ZGNR case, even though the spin stiffness in the ferromagnetic or antiferromagnetic materials is lower, it gives the Curie temperature or the N$\acute{\textrm{e}}$el temperature \cite{Padja, Essenberg, Lezaic, Jakobson, Teguh1} above the room temperature. Moreover, the spin stiffness also determines the lifetime of the spin waves before decaying into the Stoner excitations. Compared to the later, the lifetime of the spin waves in the ZGNR is shorter because of the small magnetic moment of the carbon atom at the edge. The lifetime of the spin waves has been considered by Culchac $et$ $al.$ by using the dynamical susceptibility \cite{Culchac, Culchac1,Culchac2}. 
   
Previous investigations of the spin-wave excitations under the electric field have been done by Rhim and Moon \cite{Rhim}, as well as by Culchac $et$ $al.$ \cite{Culchac2}. Using the Hubbard model, Rhim and Moon showed the decrease of the spin stiffness up to a critical value of the electric field, at which the spin stiffness starts to increase but never exceeds the spin stiffness for the non-electric-field case \cite{Rhim}. By applying the same approach, Culchac $et$ $al.$ showed that the lifetime of the spin waves can be controlled using the electric field \cite{Culchac2}. Here, we would like to investigate the response of the spin stiffness under the electric field by employing the so-called generalized Bloch theorem (GBT) by considering the ferromagnetic and antiferromagnetic configurations to establish the planar spiral.

The use of the GBT is of interest since the calculation can be efficiently performed with the primitive cell. Unlike Rhim and Moon \cite{Rhim}, we never observe the increase of the spin stiffness for the AFM configuration with the increasing electric field for the same interval. However, compared to all the previous results by using the Hubbard model, we revealed a new observation that the spiral (SP) state for the FM configuration at a very small spiral wavevector appears at a certain critical value of the electric field. Indeed, it can only be done by the GBT, which can treat a set of small spiral wavevectors close to zero. In addition, the SP state cannot be observed by introducing the doping in the ZGNR both in the FM and AFM configurations \cite{Teguh2}. Therefore, it suggests that the electric field can induce the phase transition in the ZGNR.

The appearance of the SP state is of interest for the spintronic devices. Sabirianov $et$ $al.$ used the SP model to evaluate a magnetoresistance in a domain wall for the nanowires \cite{Sabirianov}. They utilized this model to study the resistance of a domain wall for the application of magnetoelectric devices. Similar treatments with different application can also be found in Refs. \cite{Klaui, Stamps}. On the other side, the chirality of the SP formation drives the direction of a domain wall in the study of spintronics \cite{Pyatakov1, Pyatakov2, Chen}, especially for the multiferroic materials, in which the domain wall is responsible for producing the ferroelectricity \cite{Tokunaga,Wu}.
  
\section{Computational Details}
The applied GBT in this non-collinear calculation was implemented within a linear combination of pseudo-atomic orbitals (LCPAO) as basis sets and the norm-conserving pseudopotential \cite{Troullier} in the OpenMX code \cite{Openmx}. In this case, the wavefunction of the Kohn-Sham equation including a spiral wavevector $\mathbf{q}$ was described by an LCPAO \cite{Teguh}
\beq
\psi_{\nu\mathbf{k}}\left(\mathbf{r}\right)&=&\frac{1}{\sqrt{N}}\left[\sum_{n}^{N}e^{i\left(\mathbf{k}-\frac{\mathbf{q}}{2}\right)\cdot\mathbf{R}_{n}}\sum_{i\alpha}C_{\nu\mathbf{k},i\alpha}^{\uparrow}\phi_{i\alpha}\left(\mathrm{\mathbf{r}-\tau_{i}-\mathbf{R}_{n}}\right)
\left(
\begin{array}{cc}
1\\
0\end{array} 
\right)\right.\nonumber\\
& &\left.+\sum_{n}^{N}e^{i\left(\mathbf{k}+\frac{\mathbf{q}}{2}\right)\cdot\mathbf{R}_{n}}\sum_{i\alpha}C_{\nu\mathbf{k},i\alpha}^{\downarrow}\phi_{i\alpha}\left(\mathrm{\mathbf{r}-\tau_{i}-\mathbf{R}_{n}}\right)\left(
\begin{array}{cc}
0\\
1\end{array}
\right)\right],\label{lcpao}
\eeq  
where the localized orbital $\phi_{i\alpha}$ is generated by a confinement method \cite{Ozaki1, Ozaki2}. At the same time, the direction of $\mathbf{q}$ will determine at what direction the magnetic moment of the magnetic atoms 
\beq
	\mathbf{M}_{i}(t)=M_{i} \left(
\begin{array}{cc}
\cos\left(\varphi_{i}^{0}+\mathbf{q}\cdot \mathbf{R}_{i}+\omega_{\mathbf{q}} t\right)\sin\theta_{i}\\
\sin\left(\varphi_{i}^{0}+\mathbf{q}\cdot \mathbf{R}_{i}+\omega_{\mathbf{q}} t\right)\sin\theta_{i}\\
\cos\theta_{i}\end{array} 
\right). \label{moment}  
\eeq
will rotate to form a spiral configuration. The deviation of the cone angle $\theta$ from the collinear magnetic order will then create the spin-wave excitations with the magnon energy $\hbar\omega_{\mathbf{q}}$ along the $\mathbf{q}$ direction. 
\begin{figure}[h!]
\vspace{-2mm}
\quad\quad\includegraphics[scale=0.6, width =!, height =!]{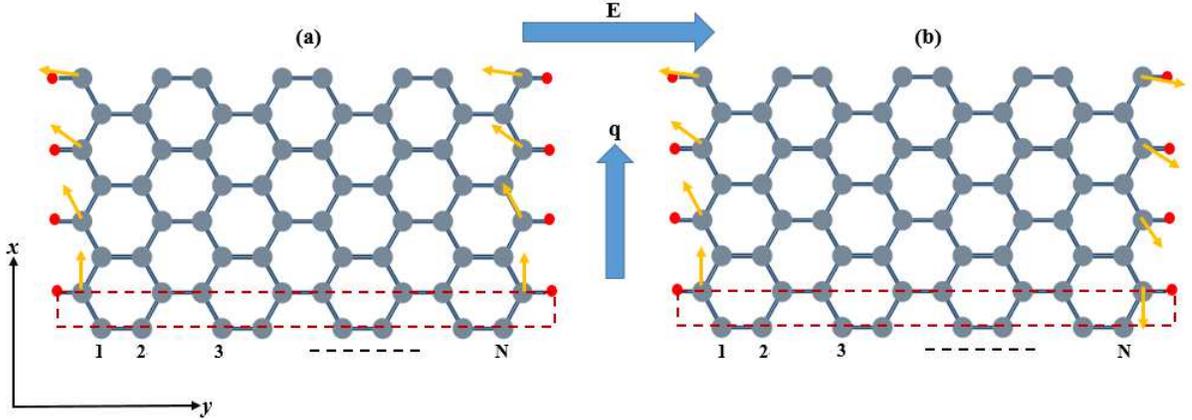}
\vspace{-2mm}
\caption{\label{model}(Color online) Configurations of planar spiral for FM (a) and AFM (b) edge states for a certain ribbon width $N$. The electric field $\mathbf{E}$ is applied along the ribbon width while the direction of $\mathbf{q}$ is parallel to $x$- axis. Here, the large gray and small red balls describe the carbon and hydrogen atoms. Moreover, the primitive unit cell is denoted by the dashed rectangle.} 
\end{figure}

The magnetic moment of the planar spiral configuration was established by setting the initial angles ($\theta_{1}=\theta_{2}=\pi/2$) and ($\varphi_{1}^{0}=\varphi_{2}^{0}=0$) for the FM configuration, and ($\theta_{1}=\theta_{2}=\pi/2$) and ($\varphi_{1}^{0}=0$, $\varphi_{2}^{0}=\pi$) for the AFM configuration, as shown in Fig. \ref{model}. We also applied the external electric field along the ribbon width of the ZGNR ($y$- axis). To avoid a deviation from the initial $\theta$, the penalty functional was employed during the self-consistent calculation to keep the planar spiral configuration \cite{Gebauer, Kurz}. By mapping the total energy difference in the self-consistent calculation onto the Heisenberg Hamiltonian, the magnon energy within the frozen magnon approach for each $\mathbf{q}$ is given by \cite{Teguh2}  
\beq 
\hbar\omega_{\mathbf{q}}=\mu_{B}\frac{E(\mathbf{q},\theta)-E(\mathbf{0},\theta)}{M}.\label{magnonf} 
\eeq   
Note that our calculation follows Edwards and Katsnelson's prediction that the spin stiffness of the ZGNR should be higher than that of 3$d$ ferromagnet metals \cite{Edwards}. 

The self-consistent calculation was carried out by using a $90 \times 1 \times 1$ $k$-point mesh, the cutoff energy of 150 Ryd, and the generalized gradient approximation (GGA) functional for treating the electron-electron interaction \cite{Perdew}. We set two $s$- and $p$- primitive orbitals for the carbon atom within the cutoff radius of 4.0 Bohr. For the hydrogen atom, two $s$- and a single $p$- primitive orbitals were used within the cutoff radius of 6.0 Bohr. For the crystal structure, the experimental lattice constant of 2.46 {\AA} for graphite was used in the \emph{x-} axis (periodic lattice) while the length of the non-periodic lattice was set to 50 {\AA} to assure the vacuum region.   

\section{Results and Discussions}
\begin{figure}[h!]
\vspace{-4mm}
\quad\quad\includegraphics[scale=0.5, width =!, height =!]{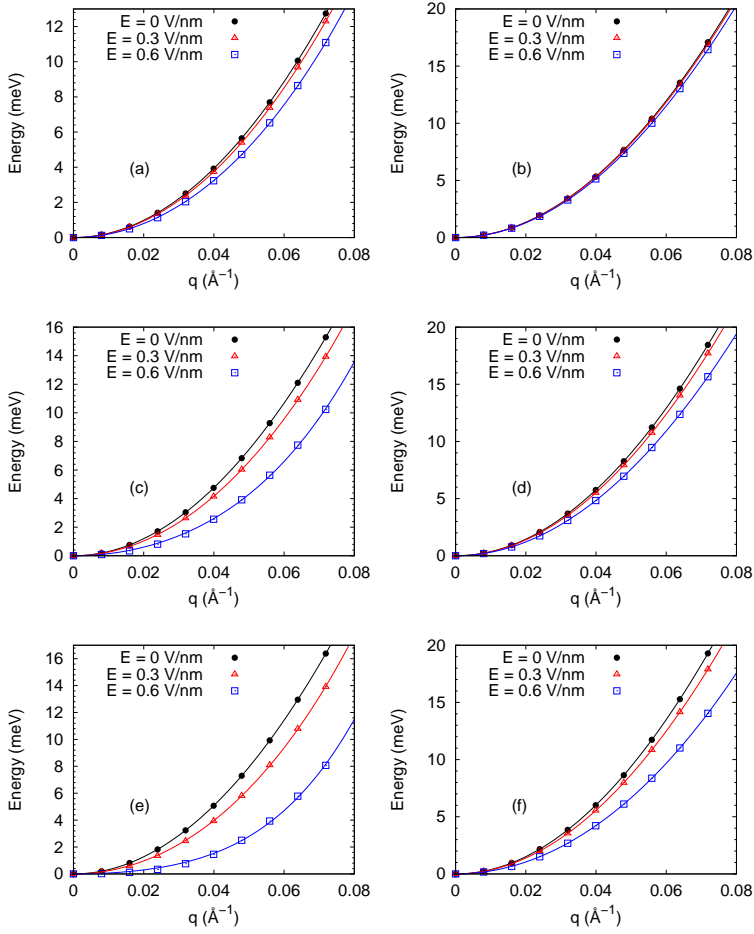}
\vspace{-2mm}
\caption{\label{stiff} (Color online) Magnon energy for FM configurations (a, c, e) and AFM configurations (b, d, f) for several values of electric field $E$. Here, the solid lines refer to the fitting function $D\mathbf{q}^{2}(1-\beta\mathbf{q}^{2})$. Figures (a, b), (c, d), and (e, f) refer to 6-ZGNR, 10-ZGNR, and 12-ZGNR, respectively.} 
\end{figure}

Before starting the discussion, here, we make an abbreviation of a ZGNR with a ribbon width $N$ as $N$-ZGNR. Then, by computing the total energy difference self-consistently for a set of $\mathbf{q}$ to calculate the magnon energy in Eq. (\ref{magnonf}), we obtain the spin stiffness $D$ by fitting the magnon energy through the relation $\hbar\omega_{\mathbf{q}}=D\mathbf{q}^{2}(1-\beta\mathbf{q}^{2})$, as shown in Fig. \ref{stiff}. Unlike Ref. \cite{Teguh2}, who used only the quadratic term of the fitting function, we include the fourth order to avoid the deviation between the data and the fitting, especially for the high electric field.  

Figure \ref{stiff} shows that the spin stiffness of the FM configuration reduces more rapidly than that of the AFM configuration as the electric field increases. Since the AFM configuration is the most stable state, the spin stiffness of the FM configuration is very sensitive to the electric field. We also notice that there is no phase transition in the AFM configuration as the electric field increases, thus showing the robustness of the ground state to the electric field. We also attempt to increase the electric field up to a critical value, at which the numerical instability appears, but no phase transition is available. On the contrary, we find an SP state for the FM configuration as the electric field increases, thus showing a phase transition from the FM state ($\textrm{q}=0$) to the SP state ($\textrm{q}>0$). This is started from a certain critical electric field, which depends on the ribbon width.
\begin{figure}[h!]
\vspace{2mm}
\quad\quad\includegraphics[scale=0.5, width =!, height =!]{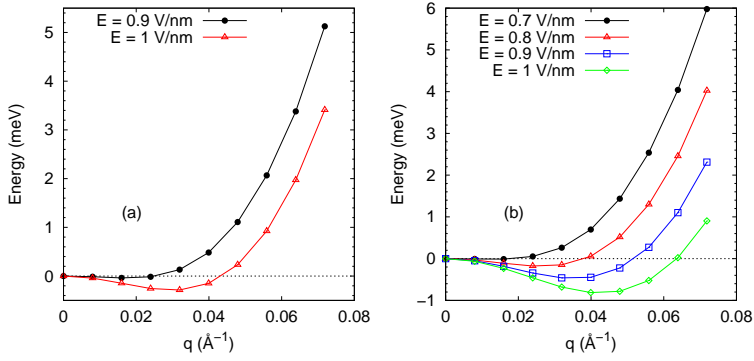}
\vspace{-2mm}
\caption{\label{spiral} Appearance of SP ground states for FM configuration in 10-ZGNR (a) and 12-ZGNR (b). The lowest value of the electric field $E$ is the critical value, at which the ground state becomes the SP state.}
\end{figure}	

\begin{figure}[h!]
\vspace{2mm}
\quad\quad\includegraphics[scale=0.5, width =!, height =!]{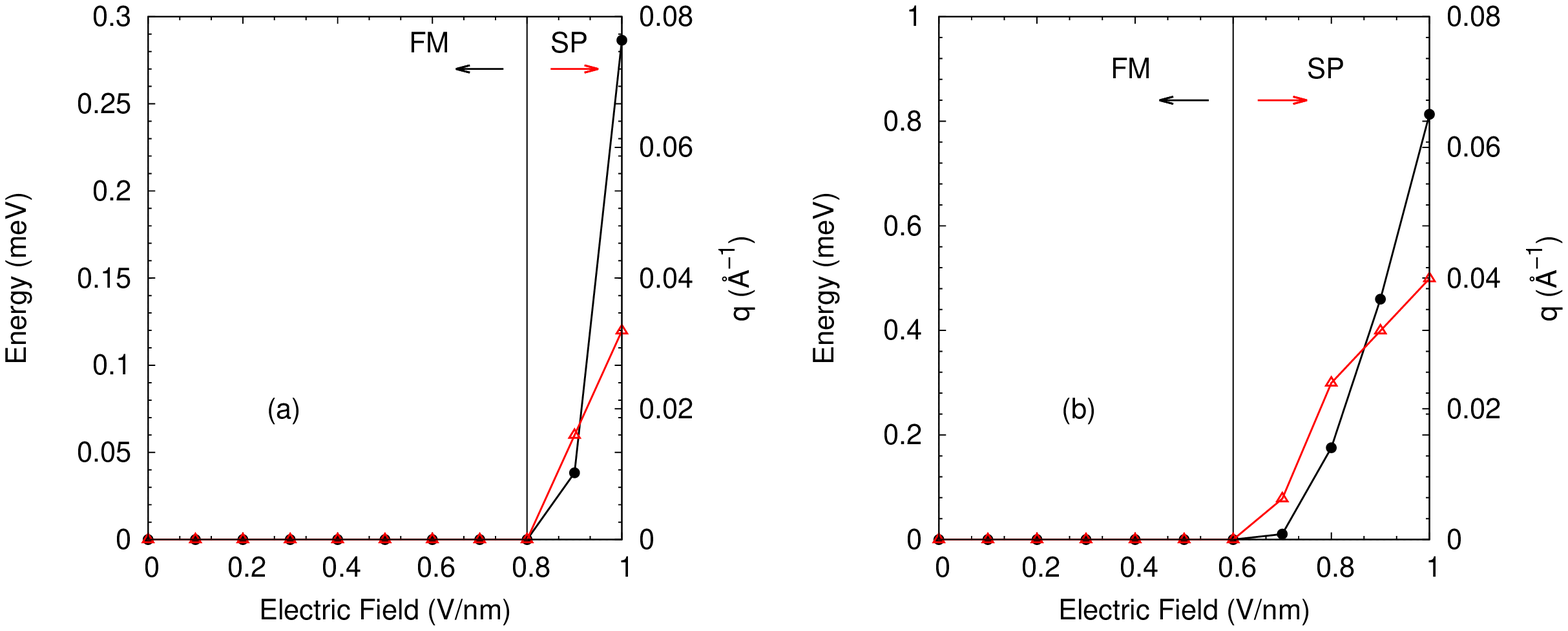}
\vspace{-2mm}
\caption{\label{phase} Phase transition from FM state to SP state in 10-ZGNR (a) and 12-ZGNR (b), respectively. Here, the black filled circles refer to the magnon energy for $\textrm{q}$, at which the energy is minimum, with respect to $\textrm{q}=0$ while the red triangles denote the appropriate $\textrm{q}$ as the most stable state.}
\end{figure}

Figure \ref{spiral} shows the SP ground state for the 10-ZGNR and the 12-ZGNR while there is no SP state for the 6-ZGNR in the interval. We see that the critical values of the electric field, from which the SP state initially appears, are 0.9 V/nm for the 10-ZGNR and 0.7 V/nm for the 12-ZGNR. Moreover, we also observe that the $\textrm{q}$, at which the energy becomes minimum, increases as the electric field increases. This creates a phase transition from the FM state to the SP state, as shown in Fig. \ref{phase}. For the 10-ZGNR in Fig. \ref{phase}(a), the interval of 0 $\leq E \leq$ 0.8 V/nm gives the FM state while $E >$ 0.8 V/nm achieves the SP state. For the 12-ZGNR in Fig. \ref{phase}(b), the critical value of $E$, at which the SP state appears, shifts to the low $E$. 0 $\leq E \leq$ 0.6 V/nm is the interval that gives the FM state while the SP state occurs in the interval of $E >$ 0.6 V/nm. So, we deduce that there is a strong dependence between the electric field and the ribbon width. We will discuss this dependence thoroughly in the next discussion.

\begin{figure}[h!]
\vspace{2mm}
\quad\quad\includegraphics[scale=0.5, width =!, height =!]{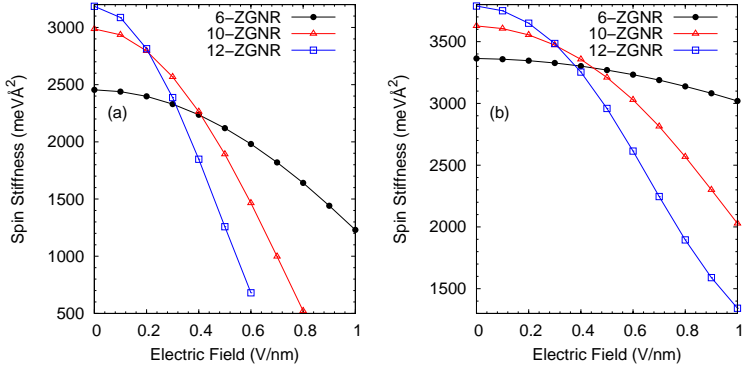}
\vspace{-2mm}
\caption{\label{tendency} Electric field dependence of spin stiffness for FM configuration (a) and AFM configuration (b) in ZGNR. Here, the ground state of the AFM configuration is robust for all the ribbon widths.}
\end{figure}

For the last result, we show the reduction of the spin stiffness as a function of the electric field, as shown in Fig. \ref{tendency}. For the FM configuration in Fig. \ref{tendency}(a), as the ribbon width increases, the critical value of $E$, which gives the SP state, decreases. In this case, the spin stiffness cannot be obtained by using the frozen magnon approach. Even though we cannot see the SP state in the 6-ZGNR for the given interval of $E$, our further calculation shows that starting from $E \geq$ 1.5 V/nm, an SP state occurs. This means that we require a sufficiently high electric field to generate the SP state for the small ribbon width, in which the exchange coupling $J_{ij}$ is stronger than that for the larger one. Thus, the stronger $J_{ij}$ can compensate the low electric field to induce the SP state in the FM configuration. On the contrary, we never observe the appearance of the SP state for the AFM configuration for all ribbon widths, as shown in Fig. \ref{tendency}(b). This indicates that the AFM configuration as the most stable configuration in the ZGNR tends to maintain its stability of the ground state. In addition, the spin stiffness decreases more rapidly as the ribbon width increases. This signifies that the small ribbon width can overcome the effect of the electric field.   

Based on the results, we would like to give our analysis for the stability of the ground state in the ZGNR with respect to the electric field. From Fig. \ref{tendency}, it is shown that at $E=0$ the spin stiffness decreases as the ribbon width decreases. It implies that the distance between the two separated carbon atoms at the edges determines how much energy needs to create the spin-waves excitations. This means that the $J_{ij}$ between those two carbon atoms decreases as the ribbon width increases, thus increasing the required energy to excite the spin waves (the increase of spin stiffness). Figure 5 also shows an opposite tendency at the low electric field and at the high electric field. At the low electric field, the decrease of spin stiffness for all ribbon widths is not so significant, thus the spin stiffness for the larger ribbon width is stronger than that for the smaller one. On the contrary, the opposite behavior appears when the high electric field is taken into account. At the high electric field, the spin stiffness for the larger ribbon width reduces much more significantly than that for the smaller one, which means that the spin stiffness for the smaller ribbon width becomes stronger under the high electric field. Therefore, we can deduce that the high electric field reduces more significantly the $J_{ij}$ for the larger ribbon width.  

When the electric field reduces the spin stiffness more significantly in the large ribbon width, it leads to the instability of the ground state for the FM configuration than that for the AFM configuration. Since the AFM configuration gives the lowest energy in the ZGNR, it is clear that the ground state of the AFM configuration is more robust than that of the FM configuration when the electrical field is applied. We can justify that the SP state can only appear for the non-stable configuration in the ZGNR when the electric field is taken into account. It can also be understood why the critical value of the electric field decreases as the ribbon width increases. This is also due to the decrease of the $J{ij}$ between the two carbon atoms as the ribbon width increases, thus the ground state of the larger ribbon width changes more rapidly.  

As shown in Fig. \ref{stiff}, the low-energy of magnon at the ZGNR gives the temperature below the room temperature, thus getting a problem to observe the property in the ZGNR-based device at the room temperature. In addition, the energy scale of the SP ground state in Fig. \ref{spiral} is even much smaller than that in a two-dimensional system \cite{Teguh3} or in a three-dimensional system \cite{Kunes}, thus also showing a subtle property. Nevertheless, there is a chance to increase the temperature close to the room temperature. According to Yazyev \cite{Yazyev1}, a very high spin stiffness should lead to the high critical temperature in the ZGNR as an $sp$-element. This can be realized by controlling the anisotropy parameter. In our case, as previously mentioned, the enhancement of spin stiffness in the ZGNR is proportional to the increase of ribbon width without applying the electric field. So, we expect that increasing the width ribbon will increase the critical temperature. Although the GBT can be used to calculate the critical temperature, however, we cannot apply it at this time. As reported in Ref. \cite{Teguh2}, the planar spiral gives a rapid reduction of the magnetic moment of C atoms at the edges both in the FM and AFM configurations whereas the calculation of critical temperature should be done in the fixed magnetic moment. 
                
In general, the change of spin stiffness under the electric field controls to the change of critical temperature. In our ZGNR case, the decrease of magnon energy/spin stiffness will also lead to the decrease of critical temperature, which causes the ZGNR worthless for spintronic devices. Although we cannot increase the spin stiffness by applying the electric field along $y$- direction, it can be possible to increase the spin stiffness with the other physical treatments, e.g., applying the electric field along the other directions. If it can be realized, we expect that the spin stiffness gives the critical temperature close to the room temperature, which provides the ZGNR useful for the graphene spintronics. Since the larger ribbon width also gives the higher spin stiffness, we can combine the ribbon width and the applied electric field to increase the critical temperature significantly.
					
\section{Conclusions} 
We show that the electric field can reduce the spin stiffness and change the ground state starting from the critical value of the electric field in the ZGNR. While the AFM configuration as the most stable state tends to maintain its ground state, the FM configuration changes its ground state from the FM state to the SP state. In addition, the reduction of spin stiffness and the position of $\textrm{q}$, at which the minimum energy is obtained, depend on the ribbon width as the electric field increases. This means that the ribbon width determines the strength of the exchange coupling $J{ij}$ between the two carbon atoms to overcome the effect of the electric field. So, we justify that the electric field can be used either to control the spin stiffness or to generate the SP state.   
\section*{Acknowledgments}
The computations were performed using a high personal computer at the Universitas Negeri Jakarta. This research is supported by a research grant ”Penelitian Unggulan Universitas 2019 No. 15/KOMP-UNJ/LPPM-UNJ/V/2019” at the Universitas Negeri Jakarta.

\end{document}